\documentstyle[prl,aps,twocolumn,epsfig]{revtex}

\begin{document}
\twocolumn[\hsize\textwidth\columnwidth\hsize\csname @twocolumnfalse\endcsname

\title{Stability of the monoclinic phase in the ferroelectric perovskite PbZr$%
_{1-x} $Ti$_x$O$_3$.}
\author{B. Noheda \thanks{%
On leave from U. Autonoma de Madrid, Spain. Correspondent author,
e-mail:noheda@bnl.gov.}, D.E. Cox and G. Shirane.}
\address{Physics Department, Brookhaven National Laboratory, Upton, NY 11973}
\author{R. Guo, B. Jones and L.E. Cross}
\address{Materials Research Laboratory, The Pennsylvania State University, University\\
Park, PA 16802}
\maketitle

\begin{abstract}
Recent structural studies of ferroelectric PbZr$_{1-x}$Ti$_{x}$O$_{3}$ (PZT)
with x= 0.48, have revealed a new monoclinic phase in the vicinity of the
morphotropic phase boundary (MPB), previously regarded as the the boundary
separating the rhombohedral and tetragonal regions of the PZT phase diagram.
In the present paper, the stability region of all three phases has been
established from high resolution synchrotron x-ray powder diffraction
measurements on a series of highly homogeneous samples with 0.42 $\leq $x$%
\leq $0.52 . At 20K the monoclinic phase is stable in the range 0.46 $\leq $x%
$\leq $ 0.51, and this range narrows as the temperature is increased. A
first-order phase transition from tetragonal to rhombohedral symmetry is
observed only for x= 0.45. The MPB, therefore, corresponds not to the
tetragonal-rhombohedral phase boundary, but instead to the boundary between
the tetragonal and {\it monoclinic} phases for 0.46 $\leq $x$\leq $ 0.51.
This result provides important insight into the close relationship between
the monoclinic phase and the striking piezoelectric properties of PZT; in
particular, investigations of poled samples have shown that the monoclinic
distortion is the origin of the unusually high piezoelectric response of PZT.
\end{abstract}

\vskip1pc]

\narrowtext

\section{Introduction}

Exceptionally striking dielectric and piezoelectric properties are found in
PbZr$_{1-x}$Ti$_{x}$O$_{3}$ (PZT), the perovskite-type oxide system which is
the basis of practically all transducers and other piezoelectric devices.
This solid solution is cubic at high temperatures but becomes slightly
distorted at lower temperatures, where it is ferroelectric. Except for a
narrow region close to PbZrO$_{3}$, the ferroelectric phase is divided in
two regions of different symmetry, rhombohedral for Zr-rich compositions and
tetragonal for Ti-rich compositions. The highest piezoelectric response in
this system is found at the nearly vertical boundary between these two
phases, at x$\simeq $ 0.47; the so-called morphotropic phase boundary(MPB),
as defined by Jaffe {\it et al.} \cite{Jaffe}. The PZT phase diagram for
compositions around the MPB is shown in Fig.1, where the open circles
represent the data of Jaffe {\it et al.} \cite{Jaffe}, which define the MPB
above room temperature. The sharpness of this line is such that a
composition fluctuation of $\Delta $x= 0.01 corresponds to a temperature
uncertainty of $\Delta $T$\simeq $ 90 K. Recently, high resolution x-ray
diffraction measurements on extremely homogenous samples by Noheda {\it et
al.} showed that an intermediate monoclinic phase exists between the
rhombohedral and tetragonal PZT phases \cite{Noheda1,Noheda2,Noheda3}. The
observation of this monoclinic phase in two different compositions, x= 0.48 
\cite{Noheda1,Noheda3} and x=0.50 \cite{Noheda2}, has allowed a preliminary
modification of the phase diagram, as shown in Fig. 1. Furthermore, the
discovery of this phase around the MPB in PZT answers many of the questions
raised by previous investigators \cite
{Shirane1,Sawaguchi,Hanh,Cross1,Misha,Du} about the 

\begin{figure}[htb]
\epsfig{width=0.80 \linewidth,figure=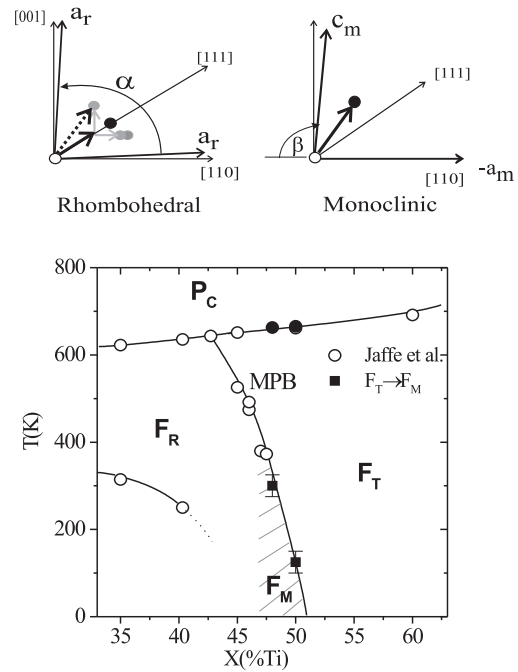}
\caption{The lower part of the figure shows the PZT phase diagram close to
the MPB reported by Jaffe et al.[1] (open symbols), and the preliminary
modification proposed in ref. 3, including the monoclinic phase. The solid
symbols represent the observed phase transitions for x= 0.48 [4] and x= 0.50
[3]. The upper part of the figure depicts the microscopic model proposed for
the rhombohedral and the monoclinic phases in ref. [4](see text).}
\label{fig1}
\end{figure}

nature of the MPB and
the underlying basis for the special physical properties of PZT in this
region of the phase diagram, especially in the context of the coexistence of
rhombohedral and tetragonal phases.

The monoclinic unit cell is doubled\ with respect to the tetragonal one and
has $b$ as the unique axis. $a_{m}$ and $b_{m}$ are directed along the
pseudo-cubic [$\overline{1}\overline{1}$0] and [1$\overline{1}$0]
directions, respectively, while c$_{m}$ is close to the tetragonal $c$ axis,
along [001], but tilted away form it such that the angle, $\beta $, between a%
$_{m}$ and c$_{m}$ is slightly larger than 90$^{\circ}$. This monoclinic
phase has unique characteristics in comparison to all other ferroelectric
perovskite phases. The polar axis is not determined by symmetry and can be
directed anywhere within the monoclinic {\it ac} plane; that is, the polar
axis is allowed to rotate within this plane. In the case of PZT, the
pseudo-cubic [111] and [001] directions are contained within the monoclinic
plane and the monoclinic polar axis is tilted away from the polar axis of
the tetragonal phase, [001], towards that of the rhombohedral phase, [111] 
\cite{Noheda3}. As has already been pointed out by other authors {\cite
{Teslic,Ricote,Corker}}, the diffraction data show clearly that the local
structure of PZT differs from that of the average one. A structure analysis
of rhombohedral PZT by Corker et al.\cite{Corker} indicated that the Zr/Ti
cations in the Zr-rich compositions are distributed among three
locally-disordered sites with monoclinic symmetry (see the gray circles in
the top-left plot of Fig. 1), resulting in average rhombohedral symmetry
(black circle in the top-left plot of Fig. 1.)\cite{Noheda3}. In a similar
structure analysis of tetragonal PZT close to the MPB\cite{Noheda3}, the
diffraction data were shown to be consistent with Zr/Ti cations distributed
among four locally-disordered cation sites with monoclinic symmetry,
resulting in average tetragonal symmetry\cite{Noheda3}.

In recent years, the development of first-principles calculations applied to
the study of ferroelectric perovskites has contributed greatly to the
understanding of the physical properties of these materials (see, e.g.,refs. 
\cite{Cohen,King-Smith,Zhong1,Garcia1,Garcia2,Rabe}). The incorporation of a
compositional degree of freedom to allow for the study of solid solutions
has been an important advance which has opened up the possibility of
investigating more complex ferroelectric materials such as PZT and related
systems \cite{Cockayne,Ghosez,Bellaiche1,Saghi,Burton,Fu,Bellaiche3}. Very recently, 
Bellaiche {\it et al}. \cite{Bellaiche4} have succeeded to derive the monoclinic phase of 
PZT from first-principles calculations. These authors also show that the value of the
piezoelectric coefficient calculated taking into account rotation of the
polarization vector in the monoclinic plane is in good agreement with the
high values observed in PZT.

In the present work, the stability region of the monoclinic phase in PZT is
characterized by means of synchrotron x-ray powder diffraction measurements
made on PZT compositions at closely-spaced intervals in the range x=
0.42-0.52. The monoclinic phase is observed at 20K for 0.46 $\leq $x$\leq $
0.51 and this composition range narrows as the temperature increases. The
transition temperature between the tetragonal and monoclinic phases is very
steep as a function of composition and coincides with the previously
mentioned MPB of Jaffe {\it et al.} \cite{Jaffe} above ambient temperature.

\begin{figure}[htb]
\epsfig{width=0.90 \linewidth,figure=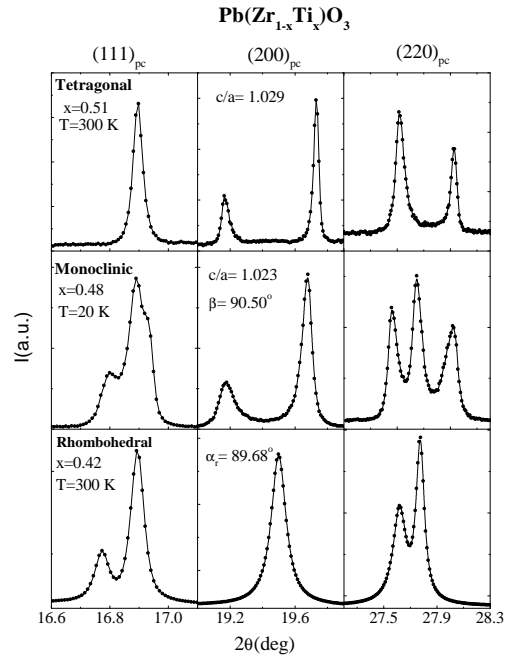}
\caption{Pseudo-cubic (111), (200) and (220) reflections for PZT with x=
0.51 at 300K (top), x= 0.48 at 20K (center) and x= 0.42 at 300 K (bottom),
showing the distinctive features of the tetragonal, monoclinic and
rhombohedral PZT phases, respectively.}
\label{fig2}
\end{figure}

\section{Experimental\label{sec:level2}\protect\bigskip}

PZT samples with x= 0.42, 0.45, 0.46, 0.47, 0.50, 0.51 and 0.52 were
prepared by conventional solid-state reaction techniques similar to those
used previously for PZT with x= 0.48 \cite{Noheda3}. During the calcination,
two steps were used. First, the desired solid solution was formed at 900 $%
^{\circ }$C using the appropriate amounts of reagent-grade powders of lead
carbonate, zirconium oxide and titanium oxide with chemical purity better
than 99.9 \%. Second, the formed product was pulverized and allowed to reach 
homogeneity by heating for six hours at 850$^{\circ }$C (lower than the temperature 
at which PbO evaporates). Pellets were then pressed using an organic binder and,
after burn out of the binder, heated to 1250 ${^{\circ }}$C at a ramp
rate of 10 ${^{\circ }}$C/min, held at this temperature in a covered
crucible for 2 hours, and cooled down to room temperature. During sintering,
PbZrO$_{3}$ was used as a lead source to maintain a PbO-rich atmosphere.

Several sets of high-resolution synchrotron x-ray powder diffraction
measurements were made on different occasions at beam line X7A at the
Brookhaven National Synchrotron Light Source. Data were collected from the
ceramic disks in symmetric flat-plate reflection geometry using $\theta $-$%
2\theta $ scans over selected angular regions in the temperature range
20-750 K. The sample was rocked 1-2${^{\circ }}$ during data collection to
improve powder averaging. In all these experiments a Ge(111) double-crystal
monochromator was use to provide an incident beam with a wavelength close to
0.7 \AA \ . A Ge(220) crystal analyzer and scintillation detector were
mounted in the diffracted beam, giving an instrumental resolution of about
0.01${^{\circ }}$ on the 2$\theta $ scale. As described in ref. \cite
{Noheda3}, measurements above room temperature were performed with the disk
mounted inside a wire-wound BN tube furnace. The accuracy of the temperature
in the furnace is estimated to be about 10 K. For low-temperature
measurements, the pellet was loaded in a closed-cycle He cryostat, which has
an estimated temperature accuracy of 2 K. With this type of diffraction
geometry it is not always possible to eliminate preferred orientation and
texture effects, but the peak positions, on which the present results are
based, are not affected.

In many cases the peak profiles were quite complex, necessitating a very
detailed and careful peak-fitting analysis. The peak positions were
determined from least-squares fits to the profile recorded for each of the
selected regions. A pseudo-Voigt peak shape function with an asymmetry
correction was used \cite{Finger}, and factors such as anisotropic peak
widths, coexisting phases and diffuse scattering between peaks were taken
into account. The lattice parameter of individual phases were obtained from
fits to the observed peak positions for several reflections. Because of the
complicated peak shapes, we found that the above procedure gave more
consistent results than standard profile-fitting programs.

Examples of selected regions of the diffraction patterns for the three PZT
phases, tetragonal (top), monoclinic (center) and rhombohedral (bottom),
around the morphotropic phase boundary are shown in Figure 2. The narrow
width of the peaks demonstrates the excellent quality of the ceramic samples
and allows the specific characteristics of each phase to be clearly
distinguished. In particular, the monoclinic phase exhibit unique features
that cannot be accounted for either of the other phase or a mixture of them.
In the monoclinic phase the unit cell is doubled in volume with respect to
the tetragonal one, with $a_{m}$ and $b_{m}$ lying along the tetragonal [$%
\overline{1}\overline{1}$0] and [1$\overline{1}$0] directions, and $c_{m}$
tilted slightly away from the [001] direction. The monoclinic phase
illustrated in this figure corresponds to the composition x= 0.48 at 20K,
described in detail in ref.\cite{Noheda3}, with $a_{m}$= 5.721 \AA , $b_{m}$%
= 5.708 \AA , $c_{m}$= 4.138 \AA and $\beta $= 90.50${^\circ}$. The $c/a$
value in Fig. 2 (center) is defined as $2\surd 2c_{m}/(a_{m}+b_{m})$, in
order to correspond to the $c_{t}/a_{t}$ ratio in the tetragonal case (top).

\begin{figure}[htb]
\epsfig{width=0.95 \linewidth,figure=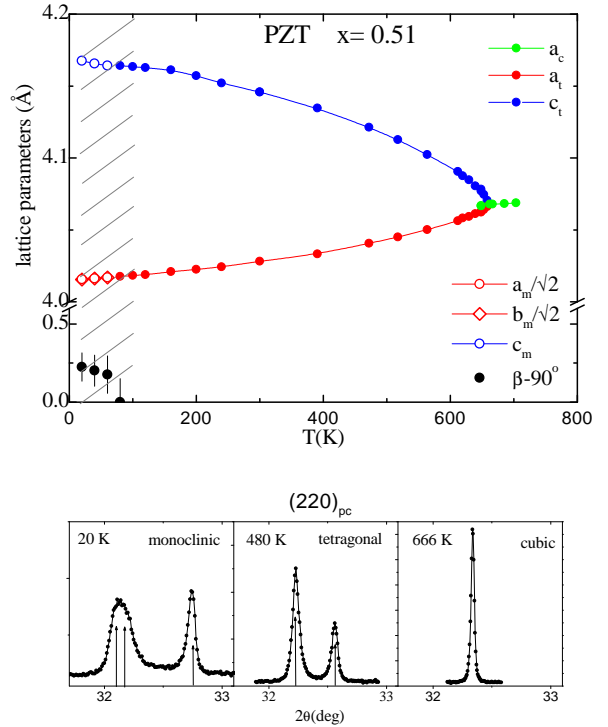}
\caption{Temperature evolution of the lattice parameters of PZT with x= 0.51
from 20 K to 700 K, for the monoclinic, $a_m$, $b_m$, $c_m$ and $\protect%
\beta $; tetragonal, $a_t$ and $c_t$; and cubic, $a_c$, phases. The dashed
region in the figure represents the uncertainty in the tetragonal-monoclinic
phase transition. At the bottom of the figure the pseudo-cubic (220)
reflection is plotted at 20 K (monoclinic), 480 K (tetragonal) and 666 K
(cubic), to illustrate the differences between the three phases.}
\label{fig3}
\end{figure}

\section{Phase transitions}

The evolution of the different structures as a function of temperature has
been determined for all the PZT samples in the present study (x= 0.42, 0.45,
0.46, 0.47, 0.51 and 0.52), and combined with previous data obtained for x=
0.48 and x= 0.50 \cite{Noheda1,Noheda3}. These results give a complete
picture of the phase transitions occurring around the morphotropic phase
boundary from 20 to 700K. Three different low temperature phases, with
rhombohedral, monoclinic and tetragonal symmetry, are observed. An important
result is that the MPB defined by Jaffe {\it et al.} \cite{Jaffe} is shown
to correspond to the limit of the {\it monoclinic} phase rather than that of
the rhombohedral phase, and is a very robust line that is reproduced for all
the samples under investigation. Both the tetragonal-monoclinic and the
tetragonal-rhombohedral phase transitions will be described in this section,
as well as the rhombohedral-rhombohedral phase transitions observed for x=
0.42 at lower temperatures.

\begin{figure}[htb]
\epsfig{width=0.90 \linewidth,figure=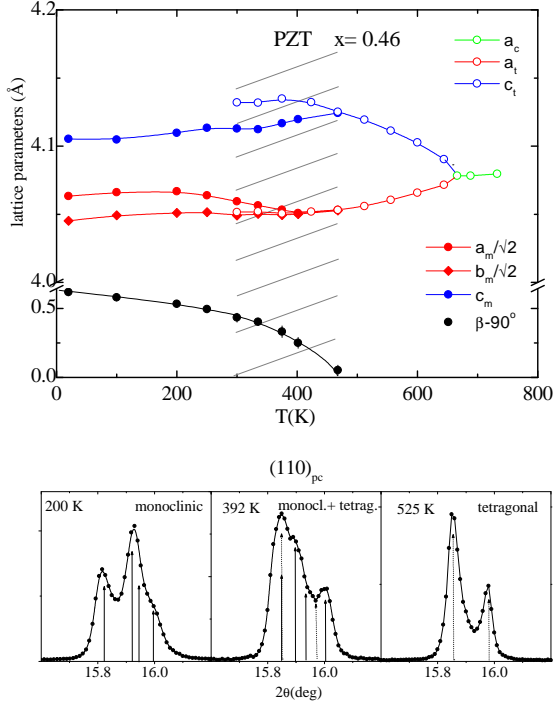}
\caption{Temperature evolution of the lattice parameters of PZT with x= 0.46
from 20 K to 710 K, for the monoclinic, $a_m$, $b_m$, $c_m$ and $\protect%
\beta $; tetragonal, $a_t$ and $c_t$; and cubic, $a_c$, phases. The dashed
area in the figure represents the region of tetragonal and monoclinic phase
coexistence. At the bottom of the figure the pseudo-cubic (110) reflection
is plotted at 200 K (monoclinic phase), 392 K (tetragonal/monoclinic phase
coexistence) and 525 K (tetragonal phase), where the calculated peak
positions are indicated by solid arrows in the monoclinic phase and dashed
arrows in the tetragonal phase.}
\label{fig4}
\end{figure}

\subsection{\protect\bigskip Tetragonal-monoclinic phase transition}

The tetragonal phase in PZT is very similar to that of pure PbTiO$_{3}$ \cite
{Glazer&Mabud,Nelmes,Shirane&Pepinski}. The effects of Zr substitution on
the structure of the tetragonal phase are basically two: first, as the Zr
content increases, the tetragonal strain, $c_{t}$/$a_{t}$, decreases, and,
second, the cubic-to-tetragonal phase transition evolves from first-order to
second-order. Figure 3 (top) shows the lattice parameters of PZT with x=
0.51 as a function of temperature. The paraelectric-ferroelectric phase
transition at T$\simeq $ 660 K is of second order, as expected \cite{Rosetti}%
, and the ferroelectric phase is purely tetragonal down to 100 K. Below this
temperature structural changes can be noticed; in particular,the tetragonal $%
(h0h)_{t}$ and $(hhh)_{t}$ reflections broaden markedly. This is apparent in
the lower part of Fig. 3, where the pseudo-cubic $(220)_{pc}$ reflections
are shown at high temperature (right), at an intermediate temperature
(center), and at low temperature (left). Based on a careful peak-fitting
analysis, the broadening at low temperatures of these reflections can be
attributed to two separated peaks, consistent with the monoclinic symmetry
observed in PZT with x=0.48 \cite{Noheda3}, also illustrated in Fig. 2.
However, the monoclinic distortion is quite small being $a_{m}$$\simeq $$%
b_{m}$ and $\beta \simeq 90.2^{\circ }$. Similar behavior was observed for a
sample of PZT with x= 0.50 \cite{Noheda2} prepared under slightly different
conditions, to be discussed later. As seen from Fig. 3, the monoclinic
angle, $\beta $, is small and the tetragonal-to-monoclinic transition
temperature can only be approximately defined at $T_{T-M}\simeq 50 $ K. On
the other hand, data collected from PZT with x= 0.52 show a well-defined
tetragonal phase down to 20K.

The evolution of the lattice parameters with temperature for PZT with x=
0.46 is shown in Figure 4 (top). The features displayed by this composition
are similar to those of PZT with x= 0.48 \cite{Noheda3}. A comparison with
the latter data at low temperatures shows that the monoclinic angle, $\beta $%
, is larger for x= 0.46 than for x= 0.48. With decreasing x (Ti content),
the differences between $a_{m}$ and $b_{m}$ also increase, while the
difference between $a_{m}$ and $c_{m}$ decreases, corresponding to the
evolution to a rhombohedral phase in which $" a_{m}=b_{m}=c_{m}"$ \cite
{Noheda3}. The monoclinic phase is very well-defined at low temperatures, as
shown by the pseudo-cubic $(110)_{pc}$ reflections plotted at the bottom
left of Fig 4. The evolution of $\beta $-$90{^{\circ }}$ shows a transition
to a tetragonal phase at $T_{T-M}\simeq 450K$, in agreement with the MPB of
Jaffe {\it et al.}. However, the characteristic features of the tetragonal
phase also appear well below this temperature, and there is a wide region of
phase coexistence between the tetragonal and monoclinic phases, as shown in
the central plot at the bottom of the figure. In this plot the peak
positions for the pseudo-cubic $(110)_{pc}$ reflections corresponding to the
monoclinic and tetragonal phases are shown together with the experimental
data. From the observed data, a reliable peak fitting analysis can be
carried out and the lattice parameters determined for both phases in this
region, as plotted as a function of temperature at the top of Fig. 4. The
measurements on PZT with x= 0.47 show similar behavior, but with a narrower
coexistence region (300 K $<$ T$<$ 400 K). For this composition the
evolution of the order parameter, $\beta -90{^{\circ }}$, suggests a
tetragonal-to-monoclinic phase transition at $T_{T-M}\simeq $ 310 K, very
close to that observed for x= 0.48, but the sample is not fully tetragonal
until the temperature is larger than 400 K, corresponding to the MPB of
Jaffe {\it el al.} for this composition.

\subsection{\protect\bigskip Tetragonal-rhombohedral phase transition}

A similar analysis for PZT with x= 0.45 yields completely different results,
as shown by the evolution of the $(200)_{pc}$ reflection, in the lower part
of Fig. 5. At low temperature the sample is rhombohedral (left) and remains
rhombohedral up to T $\simeq $ 500 K, while for T $>$ 550 K, this
composition is tetragonal. Some diffuse scattering is observed between the
tetragonal peaks, as shown in the bottom right of Fig. 5. This feature is
present in all compositions in the study, as previously noted in ref. \cite
{Noheda3}, and is associated with the existence of twin boundaries in the
tetragonal ferroelectric phase \cite{Ustinov}. From the evolution of the
order parameter (90$^{\circ }$ - $\alpha _{r}$), it is possible to determine
that the tetragonal-rhombohedral phase transition is complete at $%
T_{T-R}\simeq 580K$. A coexistence region is observed in the interval 500 K $%
<$ T$< $ 580 K. In the central plot at the bottom of the figure, the $%
(200)_{pc}$ reflection in this region is depicted, together with the
calculated peak positions for both phases.

\begin{figure}[htb]
\epsfig{width=0.90 \linewidth,figure=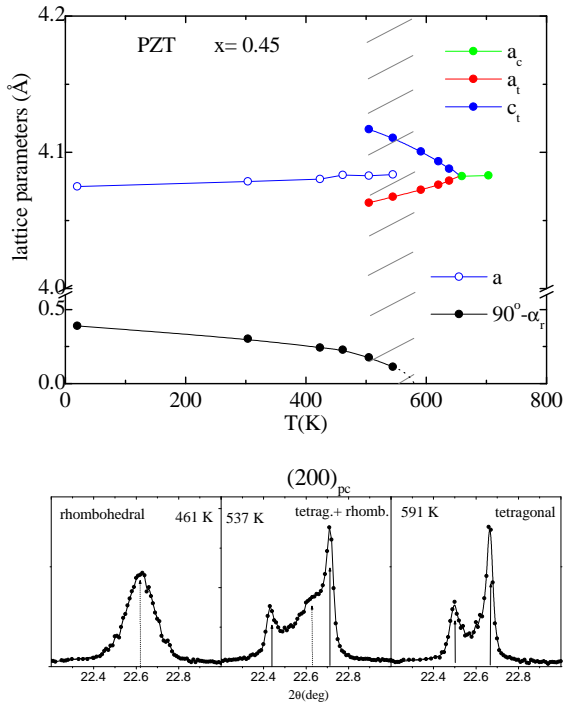}
\caption{Temperature evolution of the lattice parameters of PZT with x= 0.45
from 20 K to 710 K, for the rhombohedral, $a$ and $\protect\alpha_r$;
tetragonal, $a_t$ and $c_t$; and cubic, $a_c$, phases. The dashed area in
the figure represents the region of tetragonal and rhombohedral phase
coexistence. At the bottom of the figure the pseudo-cubic (200) reflection
is plotted at 461 K (rhombohedral phase), 537 K (tetragonal/rhombohedral
phase coexistence) and 591 K (tetragonal phase), where the calculated peak
positions are indicated by dashed arrows in the rhombohedral phase and solid
arrows in the tetragonal phase.}
\label{fig5}
\end{figure}

\begin{figure}[htb]
\epsfig{width=0.90 \linewidth,figure=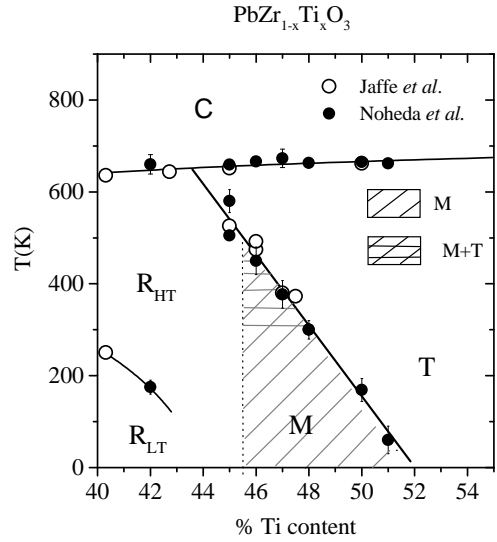}
\caption{New PZT phase diagram around the MPB. The solid symbols are the
results from the current work, together with those in ref. 3 (x= 0.50) and
ref. 4 (x= 0.48). Data from Jaffe {\it et al.} [1] and Amin {\it et al.}[36]
are represented by open circles. The monoclinic region is shaded with
diagonal lines. Horizontal lines are superimposed in the region of
tetragonal-monoclinic phase coexistence. For x= 0.45 the solid symbols
represent the limits of the tetragonal-rhombohedral coexistence region.}
\label{fig6}
\end{figure}

\subsection{Low temperature rhombohedral phase}

The data obtained for the PZT sample with x= 0.42 show that this composition
has rhombohedral symmetry all the way down to 20 K from the Curie point at $%
T_{c}$$\simeq $ 650 K. At 20 K the rhombohedral lattice parameters are $a$=
4.0921 \AA \ and $\alpha _{r}$= 89.61$^{\circ }$. With increasing
temperature, the rhombohedral angle, $\alpha _{r}$, increases gradually
until the cubic phase is reached, while $a$ remains practically constant.

Two different rhombohedral phases have been observed in PZT: a
high-temperature phase ($F_{R(HT)}$) and a low temperature rhombohedral
phase ($F_{R(LT)}$) \cite{Barnett}, which have space groups R3m and R3c,
respectively. In the latter phase, adjacent oxygen octahedra along the [111]
polar axis are rotated about this axis in opposite directions, so that the
unit cell is doubled with respect to the high-temperature phase \cite
{Michel,Glazer}. The corresponding phase boundary was also determined by
Jaffe et al.\cite{Jaffe} in the region of the phase diagram above room
temperature. An extension of this boundary below room temperature was
reported in a neutron powder diffraction study by Amin et al. \cite{Amin},
who investigated the superlattice peaks from a sample with x= 0.40 and found
the transition temperature to occur at about 250 K. In the present work, we
were also able to observe very weak superlattice peaks from a composition
with x= 0.42 below room temperature in the synchrotron x-ray patterns. The
phase boundary in this case was found to lie at approximately 175 K (see
fig. 6).

We have also observed one very weak superlattice peak in a recent neutron diffraction study 
of a sample with x= 0.48 at 20 K. This peak can be indexed in terms of a monoclinic cell with 
a doubled c-axis, but the nature of the distortion and any possible relationship with that in 
the low-temperature rhombohedral phase has not yet been determined.

\section{Discussion}

The results presented above are summarized and compared with previous data
from x= 0.48 and x= 0.50 in Figure 6, which represents the new PZT phase
diagram around the MPB. The data obtained for the
tetragonal-(monoclinic/rhombohedral) transition temperatures for 0.45$<$ x$<$%
0.51 are very consistent and lie on a well-defined line, which reproduces
the MPB of Jaffe et al. \cite{Jaffe} above ambient temperature. The boundary
between the rhombohedral and monoclinic regions is shown as a vertical line
between 0.45$< $ x$<$ 0.46, since no evidence of a monoclinic-rhombohedral
phase transition has been observed. The lattice parameters at 20 and 300 K
for the compositions under study are listed in Table 1, which also shows
clearly the widening of the monoclinic region at lower temperatures. Figure
7 shows the evolution of the lattice parameters of the different phases as a
function of composition at 300 K, from the rhombohedral to the tetragonal
PZT phases via the monoclinic phase. At
\widetext
\begin{table}[tbp]
\caption{Lattice parameters at 20 and 300K for PZT with x in the range
0.42-0.52. The symmetry, S, of the unit cell, rhombohedral(R), monoclinic
(M) or tetragonal (T) is indicated in each case. For rhombohedral symmetry $%
a=b=c$ and $\protect\alpha_r$ is the rhombohedral angle. In the monoclinic
case, $\protect\beta $ is the monoclinic angle. In the tetragonal case $%
a=b=a_{t}$ and $\protect\beta =90^{\circ }$}
\begin{tabular}{ccccccccccccc}
&  & \multicolumn{5}{c}{20 K} &  & \multicolumn{5}{c}{300 K} \\ \hline
\% Ti & \multicolumn{1}{|c}{S} & $a$ (\AA ) & $b$ (\AA ) & $c$ (\AA ) & $%
\alpha_r$ (${{}^{\circ }})$ & $\beta $ $({{}^{\circ }})$ & 
\multicolumn{1}{|c}{S} & $a$ (\AA ) & $b$ (\AA ) & $c$ (\AA ) & $\alpha_r $ $%
({{}^{\circ })}$ & $\beta $ $({{}^{\circ }})$ \\ \hline
42 & \multicolumn{1}{|c}{R} & 4.078 &  &  & 89.61 &  & \multicolumn{1}{|c}{R}
& 4.084 &  &  & 89.68 &  \\ 
45 & \multicolumn{1}{|c}{R} & 4.075 &  &  & 89.61 &  & \multicolumn{1}{|c}{R}
& 4.079 &  &  & 89.69 &  \\ 
46 & \multicolumn{1}{|c}{M} & 5.747 & 5.721 & 4.104 &  & 90.62 & 
\multicolumn{1}{|c}{M} & 5.754 & 5.731 & 4.103 &  & 90.47 \\ 
47 & \multicolumn{1}{|c}{M} & 5.731 & 5.713 & 4.123 &  & 90.58 & 
\multicolumn{1}{|c}{M} & 5.720 & 5.715 & 4.142 &  & 90.22 \\ 
48 & \multicolumn{1}{|c}{M} & 5.717 & 5.703 & 4.143 &  & 90.53 & 
\multicolumn{1}{|c}{T} & 4.041 &  & 4.140 &  &  \\ 
50 & \multicolumn{1}{|c}{M} & 5.693 & 5.690 & 4.159 &  & 90.35 & 
\multicolumn{1}{|c}{T} & 4.032 &  & 4.147 &  &  \\ 
51 & \multicolumn{1}{|c}{M} & 5.681 & 5.680 & 4.169 &  & 90.22 & 
\multicolumn{1}{|c}{T} & 4.028 &  & 4.146 &  &  \\ 
52 & \multicolumn{1}{|c}{T} & 4.009 &  & 4.158 &  &  & \multicolumn{1}{|c}{T}
& 4.030 &  & 4.145 &  & 
\end{tabular}
\end{table}

\narrowtext
the top of the figure, the unit cell
volume shows an essentially linear behavior with composition in the range
studied. The monoclinic angle, $\beta $, and lattice strain, $c/a$, at 300 K
are also plotted as a function of composition in Fig. 7, where the
rhombohedral cell with lattice parameters $a$ and $\alpha_r $ (see Table 1)
has been expressed in terms of the monoclinic cell \cite{footnote}. $c/a$
corresponds to $c_{t}/a_{t}$, $2\surd 2c_{m}/(a_{m}+b_{m})$ and $1$, in the
tetragonal, monoclinic and rhombohedral cases, respectively. $\beta $ is $%
90^{\circ }$ for a tetragonal cell and is the monoclinic angle for the
monoclinic cell. The role of the monoclinic phase as a "bridge" between the
tetragonal and rhombohedral phases in PZT is clearly demonstrated in these
plots. The monoclinic phase has also been observed by Raman scattering in a very recent 
paper by Souza Filho et al. \cite{Souza}.

The structural studies reported here, together with those in refs. [2-4],
comprise data from ten samples from two different origins spanning the
composition range 0.42$\leq $x$\leq $0.52. As mentioned earlier, only one of
these samples was inconsistent with the picture shown in Fig. 6, namely PZT
with x= 0.47 described in ref. \cite{Noheda2}. For this composition it was
found that the tetragonal phase transformed to a rhombohedral phase at low
temperatures, while, at intermediate temperatures, a poorly-defined region
of coexisting phases was observed. On the other hand, the data for the
x=0.47 sample studied in the present work shows, as described above,
characteristics similar to those of x= 0.46 or x= 0.48; in particular, the
existence of a monoclinic phase at low temperatures and no traces of a
rhombohedral phase. It is noteworthy that an analysis of the peak-widths in
the cubic phase shows clear differences in the microstructure of the two
sets of samples. The microstrain, $\Delta $d/d, and crystallite-size of the
samples used in the present study are estimated to be about 3x10$^{-4}$ and $%
1\mu $m, respectively \cite{Noheda3}. A similar analysis for the x= 0.47
sample described in ref.\cite{Noheda2} yields values of 11x$10^{-4}$ and 0.2 
$\mu $m, respectively. One possible explanation is that because of the
smaller crystallite size

\begin{figure}[htb]
\epsfig{width=0.90 \linewidth,figure=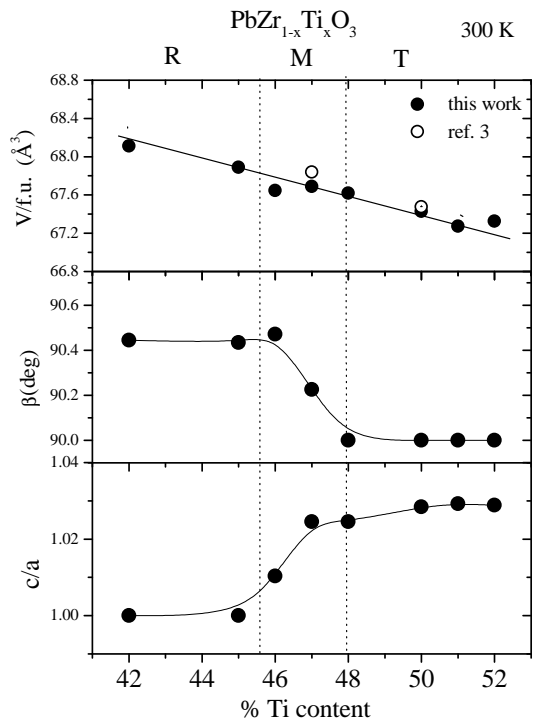}
\caption{The structural evolution with composition from the rhombohedral to
the tetragonal PZT phases, through the monoclinic phase, as illustrated by
the cell volume per formula unit, V (top), the monoclinic angle, $\protect%
\beta $ (center) \protect\cite{footnote}, and the lattice strain , $c/a$
(bottom) for PZT with 0.42$\leq $x$\leq $ 0.52 at 300 K. The cell volumes of
the samples in ref.[3] are also plotted as open circles at the top of the
figure.}
\label{fig7}
\end{figure}

 in the ceramic samples in ref.\cite{Noheda2}, the
inhomogeneous internal stress ''prematurely'' induces the tetragonal-rhombohedral phase transitions and inhibits the formation of the
intermediate monoclinic phase. With a larger crystallite size, the internal
strain is more easily relieved \cite{Buensen,Tuttle}, presumably through the
formation of non-180$^{\circ }$ domains \cite{Cahn}, and the monoclinic
phase transition is stabilized.

It is interesting to address the question of why the monoclinic phase was
not observed in any of the previous studies. One important factor is the
much superior resolution of synchrotron powder diffraction equipment
compared to that of laboratory equipment; a second is the presence of wide
regions of rhombohedral-tetragonal phase coexistence in many of these
studies, due to compositional fluctuations and/or small grain sizes
\cite
{Cross1,Ari-Gur,Fernandes,Wilkinson,Leite}, which would obscure the evidence for
monoclinic symmetry. For samples prepared by conventional ''dry''
solid-state techniques, a much narrower range of compositional fluctuations 
and large grain size can be achieved by the use of a final heat treatment at 1250 $^{\circ }$C,
as in the present case, or by the use of ''semi-wet'' methods of synthesis
and lower firing temperatures \cite{Kakegawa2,Kakegawa3,Singh,Cuneyt}.
However, perhaps the key element for clarifying the phase diagram is to
carry out the structural studies at low temperatures, as clearly
demonstrated in Fig. 6.

Very recently, experiments on poled samples by Guo et al. \cite{Guo} have
further underlined the crucial role of the monoclinic phase in PZT. These
experiments have revealed that poling induces the monoclinic distortion. The
application of an electric field causes the rotation of the polar axis and
an associated monoclinic distortion, which is retained after the field is
removed. These features are shown to be the origin of the high piezoelectric
response in PZT. It is observed that for rhombohedral PZT close to the MPB
the region of stability of the monoclinic phase increases after an electric
field is applied. The field-induced monoclinic phase is found to be
considerably wider on the Zr side of the phase diagram, at room temperature,
extending at least to a Ti content of x= 0.42. These experiments \cite{Guo}
validate the microscopic model for the MPB proposed in ref.\cite{Noheda3};
i.e. the application of a field would favor one of the local sites, which
corresponds exactly to the observed monoclinic distortion (see the dashed
arrow in the top-left plot of Fig. 1), and induces the monoclinic phase (see
the top-right plot in Fig. 1). Further studies of the poled samples are in
progress, and will be reported in a subsequent publication.

Acknowledgments

The authors are especially grateful to J.A. Gonzalo and S-E. Park, who were
collaborators in the initial stages of this investigation, for their advice
and encouragement. We also wish to thank L. Bellaiche, T. Egami, E. Salje,
B.A Tuttle, D. Vanderbilt and T. Vogt for very helpful discussions.
Financial support by the U.S. Department of Energy, Division of Materials
Sciences (contract No. DE-AC02-98CH10886) and ONR (MURI project
N00014-96-1-1173) is also acknowledged.


\begin{references}
\bibitem{Jaffe}  B. Jaffe, W.R. Cook, and H. Jaffe, {\it Piezoelectric
Ceramics} (Academic Press, London, 1971).

\bibitem{Noheda1}  B. Noheda, D.E. Cox, G. Shirane, J.A. Gonzalo, L.E.
Cross, and S-E. Park, Appl. Phys. Lett. {\bf 74}, 2059 (1999).

\bibitem{Noheda2}  B. Noheda, J.A. Gonzalo, A.C. Caballero, C. Moure, D.E.
Cox, and G. Shirane. Ferroelectrics {\bf 237}, 237 (2000).
e-print:cond-mat/9907286.

\bibitem{Noheda3}  B. Noheda, J.A. Gonzalo, L.E. Cross, R. Guo, S-E. Park,
D.E. Cox and G. Shirane, Phys. Rev. B {\bf 61}, 8687 (2000).

\bibitem{Shirane1}  G. Shirane and K. Suzuki. J. Phys. Soc. Japan {\bf 7},
333 (1952).

\bibitem{Sawaguchi}  E. Sawaguchi, J. Phys. Soc. Japan {\bf 8}, 615 (1953).

\bibitem{Hanh}  L. Hanh, K. Uchino, and S. Nomura, Jpn. J. Appl. Phys. {\bf %
17}, 637 (1978)

\bibitem{Cross1}  W. Cao and L.E. Cross, Phys. Rev. B {\bf 47}, 4825 (1993).

\bibitem{Misha}  S.K. Mishra, D. Pandey, and A. Singh, Appl. Phys. Lett. 
{\bf 69}, 1707 (1996).

\bibitem{Du}  X-h Du, J. Zheng, U. Belegundu, and K. Uchino, Appl. Phys.
Lett {\bf 72}, 2421 (1998).

\bibitem{Teslic}  S. Teslic, T. Egami, and D. Viehland, J. Phys. Chem.
Solids {\bf 57}, 1537 (1996); Ferroelectrics {\bf 194}, 271 (1997).

\bibitem{Ricote}  J. Ricote, D.L. Corker, R. W. Whatmore, S.A. Impey, A.M.
Glazer, J. Dec, K. Roleder. J. Phys.:Condens. Matter {\bf 10}, 1767 (1998).

\bibitem{Corker}  D.L. Corker, A.M. Glazer, R.W. Whatmore, A. Stallard, and
F. Fauth, J. Phys.:Condens. Matter {\bf 10}, 6251 (1998).

\bibitem{Cohen}  R. E. Cohen, Nature (London) {\bf 358}, 136 (1992).

\bibitem{King-Smith}  R.D. King-Smith and D. Vanderbilt, Phys. Rev. B {\bf %
49 }, 5828 (1994).

\bibitem{Zhong1}  W. Zhong, D. Vanderbilt, and K. Rabe, Phys. Rev. Lett. 
{\bf 73}, 1861 (1994) and Phys. Rev. B {\bf 52}, 6301 (1995).

\bibitem{Garcia1}  A. Garcia and D. Vanderbilt, Phys. Rev. B {\bf 54}, 3817
(1996).

\bibitem{Garcia2}  A. Garcia and D. Vanderbilt, Appl. Phys. Lett. {\bf 72},
2981 (1998).

\bibitem{Rabe}  U.W. Waghmare and K.M. Rabe, Phys. Rev. B {\bf 55}, 6161
(1997).

\bibitem{Cockayne}  K. M. Rabe and E. Cockayne, {\it Proceedings of the AIP
Conference} A {\bf 436}, 61 (1998). e-print: cond-mat/9804056

\bibitem{Ghosez}  Ph. Ghosez, E. Cokayne, U.V. Waghmare, and K.M. Rabe,
Phys. Rev. B {\bf 60}, 836 (1999).

\bibitem{Bellaiche1}  L. Bellaiche, J. Padilla, and D. Vanderbilt, Phys.
Rev. B {\bf 59}, 1834 (1999).

\bibitem{Saghi}  G. Saghi-Szabo, R. E. Cohen, and H. Krakauer, Phys. Rev. B 
{\bf 59}, 12771 (1999).

\bibitem{Burton}  B.P. Burton and E. Cockayne, Phys Rev. B {\bf 60}, R12542
(1999).

\bibitem{Fu}  H. Fu and R. Cohen, Nature {\bf 403}, 281 (2000).

\bibitem{Bellaiche3}  L. Bellaiche, and D. Vanderbilt, Phys. Rev. Lett. {\bf %
83}, 1347 (1999)

\bibitem{Bellaiche4}  L. Bellaiche, A. Garcia, and D. Vanderbilt, Phys. Rev.
Lett. {\bf 84}, 5427 (2000)

\bibitem{Finger}  L.W. Finger, D.E. Cox, and A.P. Jephcoat, J. Appl.
Crystallogr. {\bf 27}, 892 (1994).

\bibitem{Glazer&Mabud}  A. M. Glazer and S. A. Mabud, Acta Cryst. B {\bf 34}
, 1065 (1978).

\bibitem{Nelmes}  R. J. Nelmes and W.F. Kuhs, Solid State Commun. {\bf 54},
721 (1985).

\bibitem{Shirane&Pepinski}  G. Shirane, R. Pepinski, and B.C. Frazer, Acta
Cryst. {\bf 9}, 131 (1956).

\bibitem{Rosetti}  G.A. Rosetti, Jr. and A. Navrotsky, J. of Solid State
Chem. {\bf 144}, 188 (1999).

\bibitem{Ustinov}  A.I. Ustinov, J.-C. Niepce, C. Valot, L.A. Olikhovska,
and F. Bernard, Mater. Science Forum {\bf 321-324}, 109 (2000).

\bibitem{Barnett}  H.M. Barnett, J. Appl. Phys. {\bf 33}, 1606 (1962).

\bibitem{Michel}  C. Michel, J-M. Moreau, G.D. Achenbach, R. Gerson, and
W.J. James, Solid State Comm. {\bf 7}, 865 (1969).

\bibitem{Glazer}  A. M. Glazer , S. A. Mabud, and R. Clarke, Acta Cryst. B 
{\bf 34}, 1060 (1978).

\bibitem{Amin}  A. Amin, R.E. Newnham. L.E. Cross, and D.E. Cox, J. Solid
State Chem. {\bf 37}, 248 (1981).

\bibitem{footnote}  The rhombohedral unit cell with lattice parameters ${a} $
and $\alpha_r$, can be expressed in terms of a monoclinic unit cell as
follows: $a_{m}=2a\cos (\alpha_r/2)$, $b_{m}=2a\sin (\alpha_r /2)$ , $%
c_{m}=a $ and $\beta =180{{}^\circ}-\arccos
[(1-2sin^{2}(\alpha_r/2))/cos(\alpha_r/2)].$

\bibitem{Souza}  A.G. Souza Filho, K.C.V. Lima, A.P. Ayala, I. Guedes,
P.T.C. Freire, J. Mendes Filho, E.B. Araujo and J.A. Eiras, Phys. Rev. B 
{\bf 61}, 14283 (2000).

\bibitem{Buensen}  W.R. Buessem, L.E. Cross, and A.K. Goswami, J. Am. Ceram.
Soc. {\bf 49}, 33 (1966)

\bibitem{Tuttle}  further details: B. Tuttle and D.A. Payne, Ferroelectrics 
{\bf 37}, 603 (1981).

\bibitem{Cahn}  R.W. Cahn, Adv. Phys, {\bf 3}, 363 (1954).

\bibitem{Ari-Gur}  P. Ari-Gur and L. Benguigui, J. Phys. D: Appl. Phys. {\bf %
8}, 1856 (1975).

\bibitem{Fernandes}  J.C. Fernandes, D.A. Hall, M.R. Cockburn, and G.N.
Greaves, Nucl. Instr. and Meth. in Phys. Res. B {\bf 97}, 137 (1995).

\bibitem{Wilkinson}  A.P. Wilkinson, J. Xu, S. Pattanaik, and S.J.L.
Billinge, Chem. Mater. {\bf 10}, 3611 (1998).

\bibitem{Leite}  E. R. Leite, M. Cerqueira, L. A. Perazoli, R. S. Nasar, and
E. Longo, J. Am. Ceram. Soc. {\bf 79}, 1563 (1996).

\bibitem{Kakegawa2}  K. Kakegawa, J. Mohri, T. Takahashi, H. Yamamura, and
S. Shirasaki, Solid State Comm. {\bf 24}, 769 (1977).

\bibitem{Kakegawa3}  K. Kakegawa, J. Mohri, S. Shirasaki, and K. Takahashi,
J. Am. Ceram. Soc. {\bf 65}, 515 (1982).

\bibitem{Singh}  A.P. Singh, S.K. Mishra, D. Pandey, Ch.D. Prasad, and R.
Lal, J. of Mat. Sci. {\bf 28}, 5050 (1993).

\bibitem{Cuneyt}  A. C\"{u}neyt Ta\c{s}, J. Am. Ceram. Soc. {\bf 82}, 1582
(1999).

\bibitem{Guo}  R. Guo, L.E. Cross, S-E. Park, B. Noheda, D.E. Cox, and G.
Shirane, Phys. Rev. Lett. {\bf 84}, 5423 (2000).
\end{references}
\end{document}